\documentclass[aps,prd,preprint,nofootinbib]{revtex4}
\usepackage{amsfonts}
\usepackage{mathrsfs}
\usepackage{graphicx}
\usepackage{amsmath}
\usepackage{amssymb}
\usepackage{multirow}
\usepackage{subfigure}
\usepackage{epsfig}
\usepackage{graphicx}
\usepackage{booktabs}
\usepackage{array}
\usepackage{tabularx}
\usepackage{slashed}

\allowdisplaybreaks[4]

\graphicspath{{fig/}{dia/}} \DeclareGraphicsExtensions{.eps}
\begin{document}
\baselineskip 20pt
\title{NLO QCD corrections to $B_c$-pair production in photon-photon collision\\ [0.7cm]}

\author{Zi-Qiang Chen$^1$\footnote[2]{chenziqiang13@mails.ucas.ac.cn}, Hao Yang$^1$\footnote[3]{yanghao174@mails.ucas.ac.cn} and Cong-Feng Qiao$^{1,2}$\footnote[1]{qiaocf@ucas.ac.cn, corresponding author}}
\affiliation{$^1$ School of Physics, University of Chinese Academy of Science, Yuquan Road 19A, Beijing 10049 \\
$^2$ CAS Key Laboratory of Vacuum Physics, Beijing 100049, China}
\author{~\\}

\begin{abstract}
~\\ [-0.3cm]
The $B_c$ meson pair, including pairs of pseudoscalar states and vector states, productions in high energy photon-photon interaction are investigated at the next-to-leading order (NLO) accuracy in the nonrelativistic quantum chromodynamics (NRQCD) factorization formalism. The corresponding cross sections at the future $e^+e^-$ colliders with $\sqrt{s}=250$ GeV and $500$ GeV are evaluated.
Numerical result indicates that the inclusion of the NLO corrections shall greatly suppress the scale dependence and enhance the prediction reliability. In addition to the phenomenological meaning, the NLO QCD calculation of this process subjects to certain technical issues, which are elucidated in details and might be applicable to other relevant investigations.

\vspace {7mm} \noindent {PACS numbers: 12.38.Bx, 12.39.Jh, 14.40.Pq, 14.70.Bh}

\end{abstract}
\maketitle

\section{Introduction}
As the only heavy meson consisting of two heavy quarks with different flavors, the $B_c$ meson is of great interest in both experiment and theory.
Study of its production and decays may enrich our knowledge on the properties of double heavy meson and the nature of perturbative QCD (pQCD).
The ground state of $B_c$ meson, $B_c^+(1S)$, was discovered by CDF Collaboration \cite{Abe:1998wi,Abe:1998fb} in 1998.
And its excited state $B_c^+(2S)$ was observed by ATLAS \cite{Aad:2014laa} and CMS \cite{Sirunyan:2019osb} Collaborations in 2014 and 2019 respectively.

Due to the large masses of bottom and charm quarks, the production of heavy quark pair can be described by pQCD, while the hadronization process can be factored by using the NRQCD factorization formalism \cite{Bodwin:1994jh}.
For inclusive $B_c$ meson production, various investigations have been carried out, including the direct production through $pp$ \cite{Chang:1992jb,Chang:1994aw,Berezhnoy:1995au,Kolodziej:1995nv}, $e^+e^-$ \cite{Yang:2013vba,Zheng:2017xgj}, $\gamma\gamma$ \cite{Berezhnoy:1994bb,Kolodziej:1994uu} and $ep$ \cite{Berezhnoy:1997er,Bi:2016vbt} collisions, and the indirect production through top quark \cite{Qiao:1996rd,Sun:2010rw}, $Z$ boson \cite{Z0_Chang,Kiselev:1993iq,Qiao:2011zc,Jiang:2015jma}, $W$ boson \cite{Qiao:2011yk,Zheng:2019egj,Chen:2019xpz} and Higgs boson \cite{Jiang:2015pah} decays.

Within QCD and quantum electromagnetic dynamics (QED), the $B_c^+$ meson is produced in accompany with an additional $b\bar{c}$ pair, which is also possible to form another $b\bar{c}$ meson, namely the $B_c$-pair may exclusively produced. Generally speaking, the experiment measurement of exclusive process possesses a relative high precision, which is required in exploring the properties of QCD and hadrons. In the literature, various $B_c$-pair production processes have been investigated, including in $pp$ \cite{Baranov:1997wy,Li:2009ug}, $e^+e^-$ \cite{Kiselev:1993iu,Karyasov:2016hfm,Berezhnoy:2016etd} and $\gamma\gamma$ \cite{Baranov:1997wy} collisions. We notice that in Ref.\cite{Baranov:1997wy} the leading order (LO) analysis on $B_c$-pair production in photon-photon collision was performed, however with only $B_c^++B_c^-$ (pseudoscalar-pseudoscalar, PP) and $B_c^{*+}+B_c^{*-}$ (vector-vector, VV) configurations being considered. In this work, for the sake of completeness we first repeat the LO calculation in \cite{Baranov:1997wy} and then calculate the LO $B_c$-pair production in $B_c^++B_c^{*-}$ (pseudoscalar-vector, PV) and $B_c^{*+}+B_c^-$ (vector-pseudoscalar, VP) configurations\footnote{The PV and VP production are related by a charge-conjugation transformation. Their cross section are exactly the same.}. In the end, all these processes will be evaluated  up to the NLO QCD accuracy.
Note, hereafter for simplicity the $B_c$ represents for both pseudoscalar $B_c$ and vector $B_c^*$, the latter may overwhelmingly decay to the pseudoscalar state, unless specifically mentioned.

The rest of the paper is organized as follows. In section \uppercase\expandafter{\romannumeral2} we present the primary formulae employed in the calculation. In section \uppercase\expandafter{\romannumeral3}, some technical details in the analytical calculation are given.
In section \uppercase\expandafter{\romannumeral4}, the numerical evaluation for concerned processes is performed.
The last section is remained for summary.

\section{Formulation}
According to NRQCD factorization formalism, the cross section of $B_c$-pair production via photon-photon fusion can be formulated as
\begin{align}
d\hat{\sigma}(\gamma+\gamma\to B_c^++B_c^-)=\frac{|\psi(0)|^4}{2\hat{s}}\frac{1}{4}\sum |\mathcal{M}(\gamma+\gamma\to [c\bar{b}]+[b\bar{c}])|^2d{\rm PS}_2\ ,
\label{eq_phphformula}
\end{align}
where $\psi(0)$ is the wave function of $B_c$ meson at the origin, $\hat{s}$ is the center-of-mass energy square of two colliding photons, $\sum$ sums over the polarizations and colors of the initial and final particles, $\frac{1}{4}$ comes from the spin average of the initial $\gamma\gamma$ states, $\mathcal{M}(\gamma+\gamma\to [c\bar{b}]+[b\bar{c}])$ is the corresponding partonic amplitude, $d{\rm PS}_2$ stands for the two-body phase space.

 The partonic amplitude can be computed by using the covariant projection operator method.
At the leading order of the relative velocity expansion, it is legitimate to take $m^{}_{B_c}=m_b+m_c$, $p^{}_{B_c}=p_c+p_b=(1+\frac{m_c}{m_b})p_b$. The spin and color projection operator has the form
\begin{equation}
\Pi(n)=\frac{1}{2\sqrt{m^{}_{B_c}}}\epsilon(n)(\slashed p^{}_{B_c}+m^{}_{B_c})\otimes\left(\frac{1_c}{\sqrt{N_c}}\right),
\end{equation}
where $\epsilon(^1S_0)=\gamma_5$, $\epsilon(^3S_1)=\slashed\epsilon$, and $\epsilon$ represents the polarization vector of $B_c^*$ meson.
The $1_c$ stands for the unit color matrix, and $N_c=3$ for the number of colors in QCD.

The photon-photon scattering may be achieved in high energy $e^+e^-$ collider like the Large Electron-Positron Collider (LEP), the Circular Electron Positron Collider (CEPC) and the International Linear Collider (ILC), or even in hadron collider like the Large Hadron Collider (LHC). Here we focus only on the $e^+e^-$ collision case, where the initial photon can be generated by the bremsstrahlung or by the laser back scattering (LBS) effect.
The cross section are then formulated as
\begin{equation}
d\sigma(e^++e^-\to e^++e^-+B_c^++B_c^-) =\int dx_1dx_2f_\gamma(x_1)f_\gamma(x_2)d\sigma(\gamma+\gamma\to B_c^++B_c^-),
\end{equation}
where $f_\gamma(x)$ is the photon distribution with fraction $x$ of the beam energy.

Imposing transverse momentum cut $p_T^-<p_T<p_T^+$ and rapidity cut $|y|<y_c$ on each $B_c$ meson, the formula for total cross section is then
\begin{align}
&\sigma(e^++e^-\to e^++e^-+B_c^++B_c^-) \nonumber \\
=&\frac{1}{256\pi}\Bigg\{\theta\big(-\ln\tfrac{2m_T^-}{\sqrt{s}}\big)
\int^{{\rm min}\{0,\ln\frac{2m_T^+}{\sqrt{s}}\}}_{\ln\frac{2m_T^-}{\sqrt{s}}}dX\int^{\min\{y_T^+,y_c\}}_{\max\{-y_T^+,-y_c\}}dy^*\frac{{\rm sech}^2y^*}{E_1^2}\sum |\mathcal{M}|^2 \nonumber \\
&\int_{\max\{-y_c+y^*,X\}}^{\min\{y_c-y^*,-X\}}dy_0 x_1f_\gamma(x_1)x_2f_\gamma(x_2)+\theta\big(-\ln\tfrac{2m_T^+}{\sqrt{s}}\big)\int^{\min\{0,\ln(\frac{2m_T^+}{\sqrt{s}}{\rm cosh}y_c)\}}_{\ln\frac{2m_T^+}{\sqrt{s}}}dX \nonumber \\
&\bigg(\int_{y_T^-}^{\min\{y_T^+,y_c\}}dy^*+\int^{-y_T^-}_{\max\{-y_T^+,-y_c\}}dy^*\bigg)\frac{{\rm sech}^2y^*}{E_1^2}\sum |\mathcal{M}|^2 \nonumber \\
&\int_{\max\{-y_c+y^*,X\}}^{\min\{y_c-y^*,-X\}}dy_0 x_1f_\gamma(x_1)x_2f_\gamma(x_2) \Bigg\}\ ,
\label{eq_totcross}
\end{align}
with
\begin{align}
&X=\frac{1}{2}\ln(x_1x_2)\ ,\quad y_0=\frac{1}{2}\ln\frac{x_1}{x_2}\ , \nonumber \\
&m_T^\pm=\sqrt{m^2_{B_c}+p_T^{\pm 2}}\ ,\nonumber \\
&y_T^\pm=\frac{1}{2}\ln\frac{E_1+\sqrt{E_1^2-m_T^{\mp 2}}}{E_1-\sqrt{E_1^2-m_T^{\mp 2}}}\ .
\end{align}
Here, $\sqrt{s}$ is the collision energy of $e^+e^-$ collider, $E_1=\sqrt{sx_1x_2}/2$ and $y^*=y-y_0$ are respectively the energy and rapidity of $B_c$ meson in the  photon-photon center-of-mass system, $\theta(x)$ means the unit step function.

\begin{figure}
	\centering
	\includegraphics[width=0.5\textwidth]{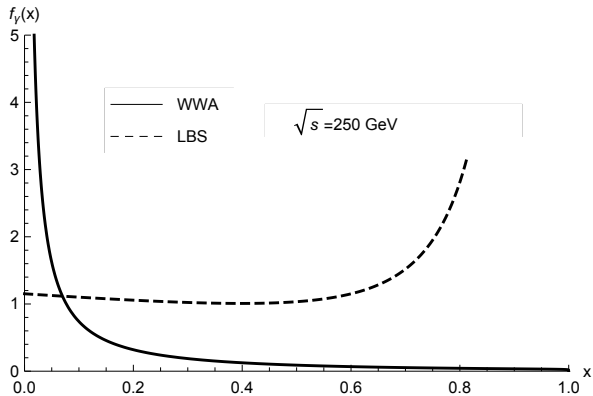}
		\caption{The spectra of WWA photon and LBS photon at $\sqrt{s}=250$ GeV.}
	\label{fig_phdis}
\end{figure}

The spectrum of  bremsstrahlung photon is well formulated in the Weizsacker-Williams approximation (WWA) as \cite{Frixione:1993yw}
\begin{equation}
f_\gamma(x)=\frac{\alpha}{2\pi}\left[\frac{1+(1-x)^2}{x}\log\left(\frac{Q^2_{\rm max}}{Q^2_{\rm min}}\right)+2m_e^2x\left(\frac{1}{Q^2_{\rm max}}-\frac{1}{Q^2_{\rm min}}\right)\right],
\end{equation}
where $Q^2_{\rm min}=m_e^2x^2/(1-x)$ and $Q^2_{\rm max}=Q^2_{\rm min}+(\theta_c\sqrt{s}/2)^2(1-x)$ with  $x=E_\gamma/E_e$, $\theta_c$ is the experimental angular cut which taken to be 32 mrad here.
For the LBS photon, the spectrum is expressed as \cite{Ginzburg:1981vm}
\begin{equation}
f_\gamma(x)=\frac{1}{N}\left[1-x+\frac{1}{1-x}-4r(1-r)\right],
\end{equation}
where $r=\frac{x}{x_m(1-x)}$ and the normalization factor
\begin{equation}
N=\left(1-\frac{4}{x_m}-\frac{8}{x_m^2}\right)\log(1+x_m)+\frac{1}{2}+\frac{8}{x_m}-\frac{1}{2(1+x_m)^2}\ .
\end{equation}
Here $x_m\simeq 4.83$ \cite{Telnov:1989sd} and the energy fraction $x$ of photon is restricted in $0\le x\le x_m/(1+x_m)$.
The behaviors of WWA photon and LBS photon are quite different, their spectra at $\sqrt{s}=250$ GeV are shown in Fig.\ref{fig_phdis}.

\section{Analytical calculation}
\begin{figure}
	\centering
	\includegraphics[width=\textwidth]{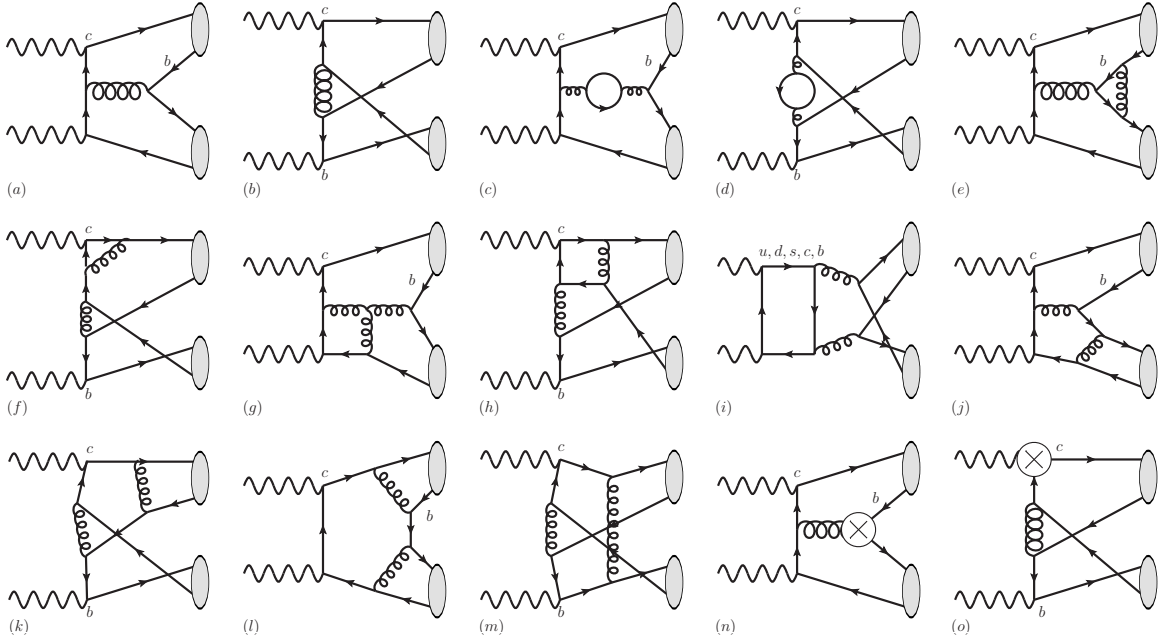}
		\caption{Typical Feynman diagrams of the partonic processes for $B_c$-pair production. (a) and (b): tree-level diagrams; (c) and (d): self-energy corrections; (e) and (f): vertex corrections; (g)-(i): box diagrams; (j) and (k): pentagon diagrams; (l) and (m): hexagon diagrams; (n) and (o): diagrams of counter terms.}
	\label{fig_FeynDia}
\end{figure}

The typical tree-level and one-loop Feynman diagrams for the partonic processes are shown in Fig.\ref{fig_FeynDia}.
The momenta and the polarization vectors of incoming and outgoing particles are denoted as:
\begin{equation}
\gamma(p_1,\epsilon_1)+\gamma(p_2,\epsilon_2)\to [c\bar{b}](k_1,\epsilon_3)+[\bar{c}b](k_2,\epsilon_4)\ .
\label{eq_partonicprocess}
\end{equation}
Here, initial and final state particles are all on their mass shells: $p_1^2=p_2^2=0$ and $k_1^2=k_2^2=m^{2}_{B_c}$.
The polarization vectors satisfy the constraints: $\epsilon_1\cdot \epsilon_1^*=\epsilon_2\cdot \epsilon_2^*=\epsilon_3\cdot \epsilon_3^*=\epsilon_4\cdot \epsilon_4^*=-1$ and $p_1\cdot \epsilon_1=p_2\cdot \epsilon_2=k_1\cdot \epsilon_3=k_2\cdot \epsilon_4=0$.

To proceed the calculation, we notice working in the photon-photon center-of-mass system is convenient. By introducing the orthonormal four-vector base: $n_0=(1,0,0,0)$, $n_1=(0,1,0,0)$, $n_2=(0,0,1,0)$ and $n_3=(0,0,0,1)$, we may choose
\begin{align}
&p_1=E_1(n_0+n_3)\ , \quad p_2=E_1(n_0-n_3)\ , \nonumber \\
&k_1=E_1(n_0+r_yn_2+r_zn_3)\ , \quad k_2=E_1(n_0-r_yn_2-r_zn_3)\ .
\label{eq_momchoice}
\end{align}
and
\begin{align}
&\epsilon_1^{(1)}=n_1\ ,\ \epsilon_1^{(2)}=n_2\ ,\ \epsilon_2^{(1)}=n_1\ ,\ \epsilon_2^{(2)}=n_2\ ; \nonumber \\
&\epsilon_3^{(1)}=n_1\ ,\ \epsilon_3^{(2)}=\frac{r_zn_2-r_yn_3}{\sqrt{r_y^2+r_z^2}}\ ,\ \epsilon_3^{(3)}=\frac{(r_y^2+r_z^2)n_0+r_yn_2+r_zn_3}{r_m\sqrt{r_y^2+r_z^2}}\ ; \nonumber \\
&\epsilon_4^{(1)}=n_1\ ,\ \epsilon_4^{(2)}=\frac{r_zn_2-r_yn_3}{\sqrt{r_y^2+r_z^2}}\ ,\ \epsilon_4^{(3)}=\frac{(r_y^2+r_z^2)n_0-r_yn_2-r_zn_3}{r_m\sqrt{r_y^2+r_z^2}}\ .
\label{eq_polchoice}
\end{align}
Here, $E_1=\sqrt{sx_1x_2}/2$, $r_y=k_y/E_1$, $r_z=k_z/E_1$, $r_m=m^{}_{B_c}/E_1$, and the on-shell condition constrains $r_y^2+r_z^2+r_m^2=1$.
Then, the helicity amplitudes can be readily calculated through
\begin{align}
&\mathcal{M}^{ij}_{\rm PP}=A_{\rm PP}^{\mu\nu}\epsilon^{(i)}_{1\mu}\epsilon^{(j)}_{2\nu}\ , \nonumber \\
&\mathcal{M}^{ijk}_{\rm PV}=A_{\rm PV}^{\mu\nu\rho}\epsilon^{(i)}_{1\mu}\epsilon^{(j)}_{2\nu}\epsilon^{(k)}_{3\rho}\ , \nonumber \\
&\mathcal{M}^{ijk}_{\rm VP}=A_{\rm VP}^{\mu\nu\rho}\epsilon^{(i)}_{1\mu}\epsilon^{(j)}_{2\nu}\epsilon^{(k)}_{4\rho}\ , \nonumber \\
&\mathcal{M}^{ijkl}_{\rm VV}=A_{\rm VV}^{\mu\nu\rho\sigma}\epsilon^{(i)}_{1\mu}\epsilon^{(j)}_{2\nu}\epsilon^{(k)}_{3\rho}\epsilon^{(l)}_{4\sigma}\ .
\end{align}

The tree-level calculation is straightforward, however the full analytic expressions of helicity amplitudes are still too lengthy to present in the mainbody of text. Considering of the symmetric property in amplitudes, we present the LO results in Appendix.

In the computation of one-loop amplitudes, the conventional dimensional regularization with $D=4-2\epsilon$ is adopted to regularize the ultraviolet (UV) and infrared (IR) singularities.
The IR singularities are canceled each other and the UV singularities are removed by renormalization procedure.
The renormalization constants include $Z_2$, $Z_m$, $Z_3$ and $Z_g$, corresponding to heavy quark field, heavy quark mass, gluon field and strong coupling constant, respectively.
We define $Z_2$ and $Z_m$ in the on-shell (OS) scheme, $Z_3$ and $Z_g$ in the modified minimal-subtraction ($\overline{\rm MS}$) scheme. The corresponding counterterms are
\begin{align}
\delta Z_2^{\rm OS}=&-C_F\frac{\alpha_s}{4\pi}
\left[\frac{1}{\epsilon_{\rm UV}}+\frac{2}{\epsilon_{\rm IR}}
-3\gamma_E+3\ln\frac{4\pi\mu^2}{m^2}+4\right],
\nonumber\\
 \delta Z_m^{\rm OS}=&-3C_F\frac{\alpha_s}{4\pi}
\left[\frac{1}{\epsilon_{\rm
UV}}-\gamma_E+\ln\frac{4\pi\mu^2}{m^2} +\frac{4}{3}\right], \nonumber\\
\delta Z_3^{\overline{\rm MS}}=&\dfrac{\alpha_s}{4\pi}(\beta_0-2C_A)\left[\dfrac{1}{\epsilon_{\rm UV}} -\gamma_E +\ln(4\pi) \right],
 \nonumber\\
  \delta Z_g^{\overline{\rm MS}}=&-\frac{\beta_0}{2}\,
  \frac{\alpha_s}{4\pi}
  \left[\frac{1}{\epsilon_{\rm UV}} -\gamma_E + \ln(4\pi)
  \right].
\end{align}
Here, $\mu$ is the renormalization scale, $\gamma_E$ is the Euler's constant; $m$ stands for $m_c$ and $m_b$ accordingly; $\beta_0=(11/3)C_A-(4/3)T_fn_f$ is the one-loop coefficient of QCD beta function, $n_f$ is the number of active quarks which taken to be 5 in our calculation; $C_A=3$, $C_F=4/3$ and $T_F=1/2$ are normal color factors.
Note, in final results, all $\delta Z_3$ terms cancel with each other.

For reference, we provide the analytic results for one-loop amplitudes as supplementary files attached to the arXiv preprint.
In the NLO calculation, the Mathematica package FeynArts \cite{Hahn:2000kx} is used to generate Feynman diagrams and Feynman amplitudes; FeynCalc \cite{Mertig:1990an,Shtabovenko:2016sxi} and FORM \cite{Vermaseren:2008kw,Kuipers:2012rf} are used to perform algebraic calculation; The package FIRE \cite{Smirnov:2008iw,Smirnov:2014hma} is employed to reduce the Feynman integrals to typical master integrals $A_0$, $B_0$, $C_0$ and $D_0$, which are numerically evaluated by LoopTools \cite{Hahn:1998yk}.

\section{Numerical results}
In numerical analysis, the formula (\ref{eq_totcross}) is employed with $|\mathcal{M}|^2\simeq |\mathcal{M}_{\rm tree}|^2$ for the LO calculation and $|\mathcal{M}|^2\simeq |\mathcal{M}_{\rm tree}|^2+2{\rm Re}(\mathcal{M}_{\rm loop}\mathcal{M}^*_{\rm tree})$ for the NLO calculation.
The rapidity and $p_T$ cuts, $|y|<2$ and $2<p_T<40$ GeV, are imposed on each $B_c$ meson.
Other inputs in numerical evaluation go as follows:
\begin{align}
&\alpha=1/137.065,\quad m_e=0.511\ {\rm Mev},\quad m_c=1.5\ {\rm GeV},\nonumber \\
&\quad\quad m_b=4.8\ {\rm GeV},\quad |\psi(0)|^2=0.174\ {\rm GeV}^3.
\end{align}
Here, the $B_c$ wave function at the origin is estimated from the $^3S_1-{}^1S_0$ splitting \cite{Eichten:1994gt}
\begin{equation}
|\psi(0)|^2=\frac{9m_bm_c}{21\pi \alpha_s}(M_{B_c^*}-M_{B_c})
\end{equation}
with the lattice calculation result on $M_{B_c^*}-M_{B_c}= 53$ MeV \cite{Gregory:2010gm}.

The two-loop strong coupling of
\begin{equation}
\frac{\alpha_s(\mu)}{4\pi}=\frac{1}{\beta_0L}-\frac{\beta_1\ln L}{\beta_0^3L^2}
\label{eq_alphasLambda}
\end{equation}
is employed in the NLO calculation, in which, $L=\ln(\mu^2/\Lambda_{\rm QCD}^2)$, $\beta_1=(34/3)C_A^2-4C_FT_Fn_f-(20/3)C_AT_Fn_f$,
with $n_f=5$ and $\Lambda_{\rm QCD}=210\ {\rm MeV}$ adopted here \cite{Tanabashi:2018oca}.
Note, for LO calculation, the one-loop formula of the running coupling constant is used.

\begin{figure}[!htbp]
\centering
\subfigure[]{
\includegraphics[width=0.46\textwidth]{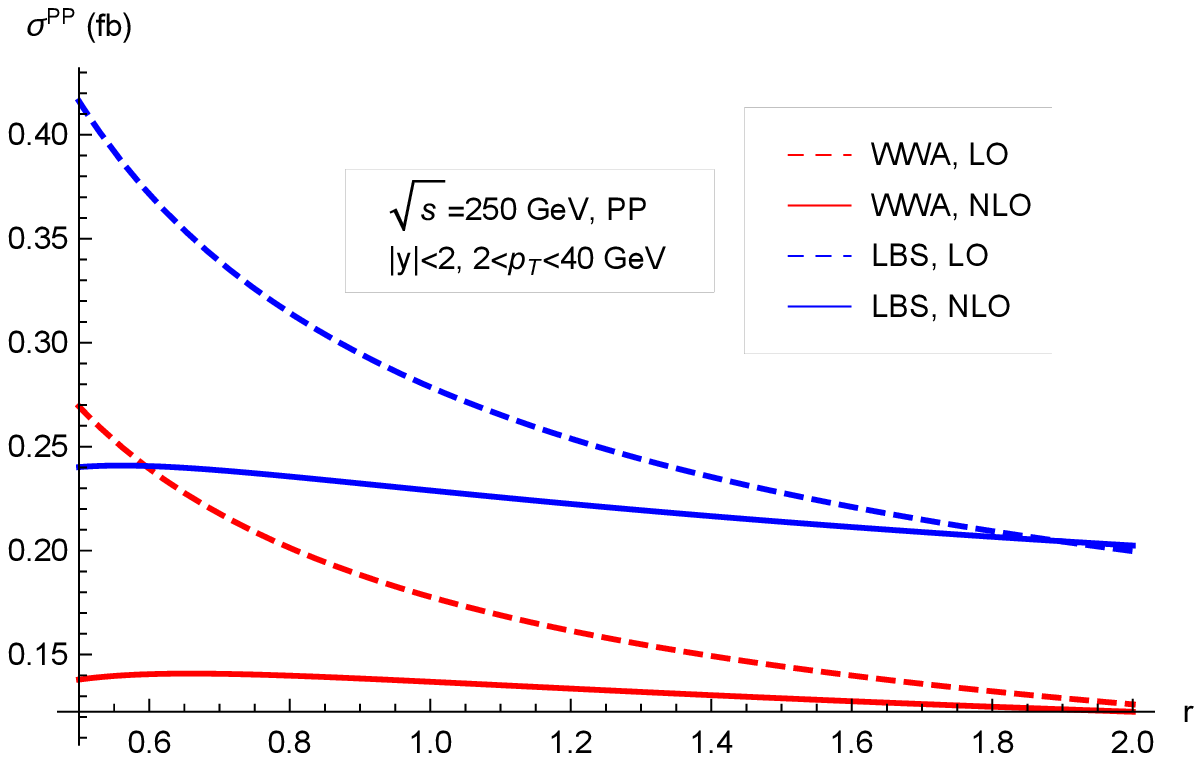}}
\subfigure[]{
\includegraphics[width=0.46\textwidth]{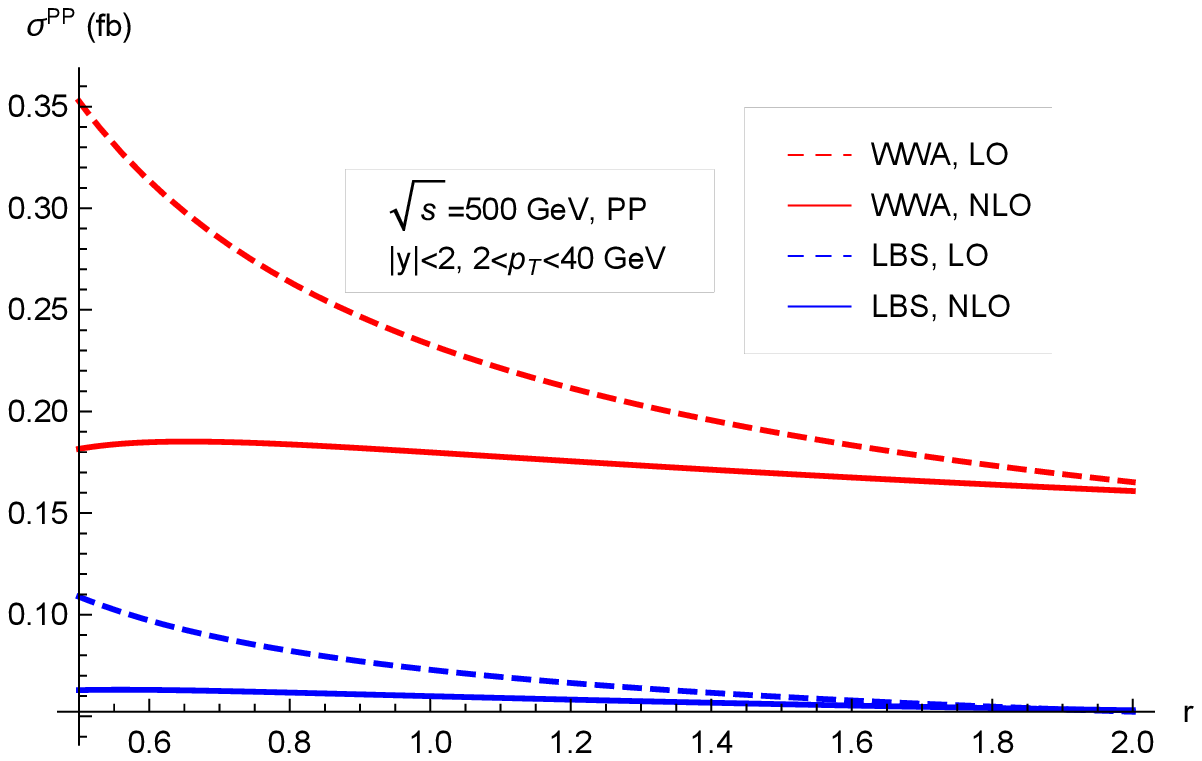}}
\subfigure[]{
\includegraphics[width=0.46\textwidth]{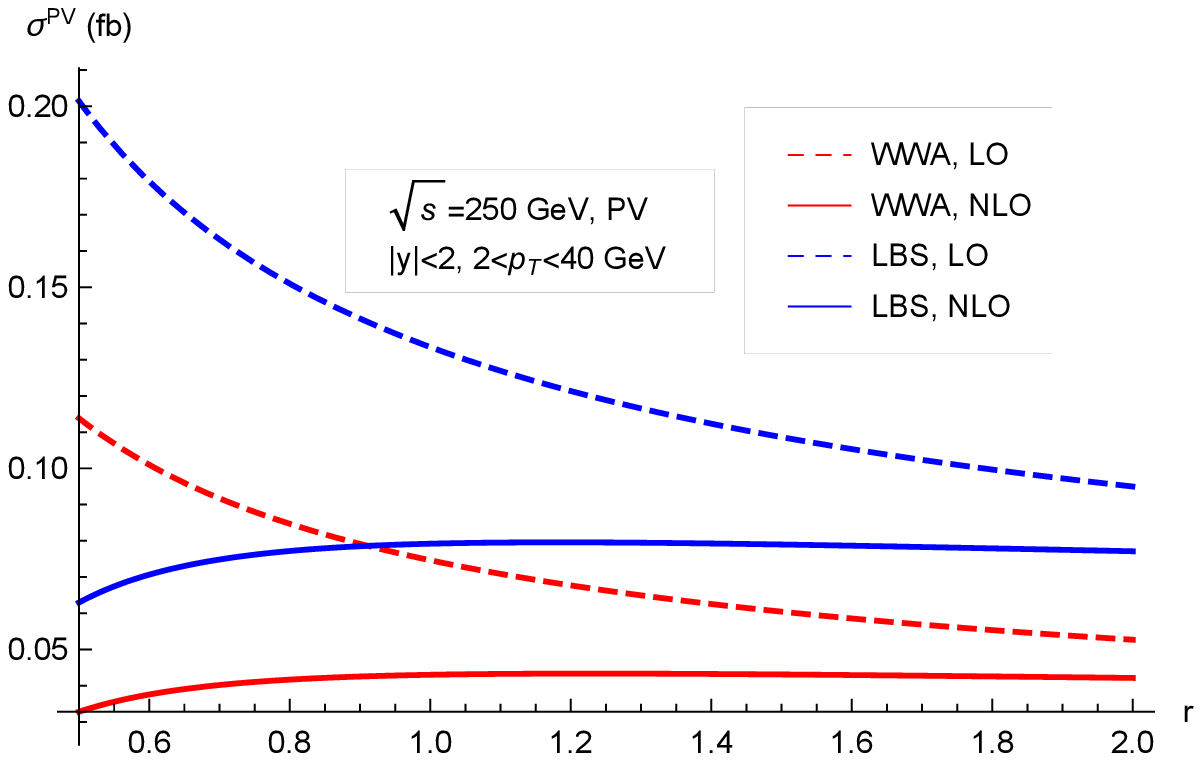}}
\subfigure[]{
\includegraphics[width=0.46\textwidth]{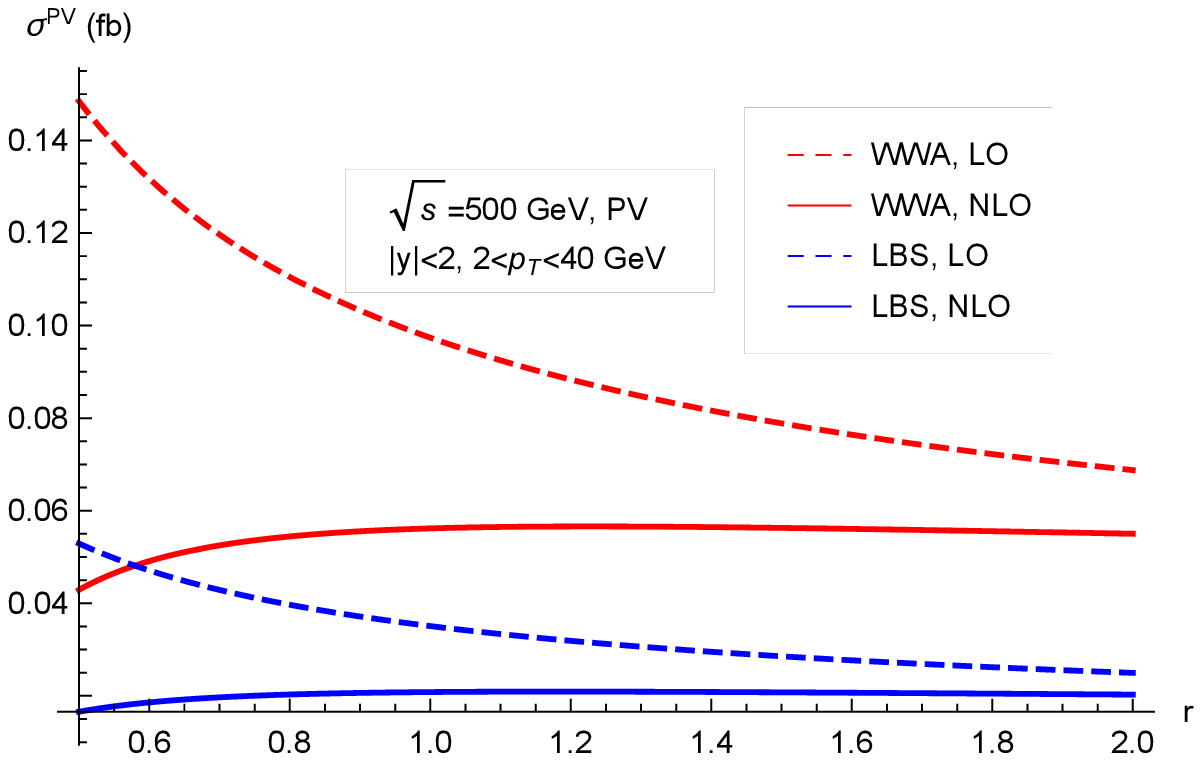}}
\subfigure[]{
\includegraphics[width=0.46\textwidth]{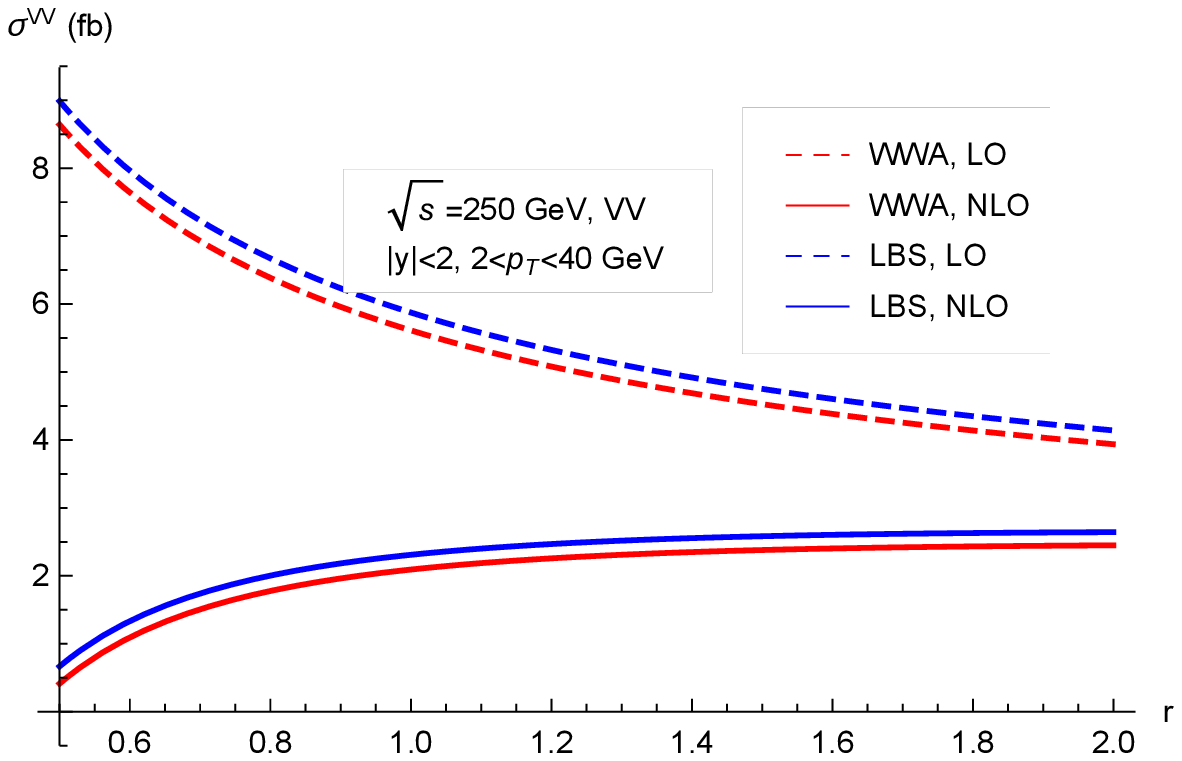}}
\subfigure[]{
\includegraphics[width=0.46\textwidth]{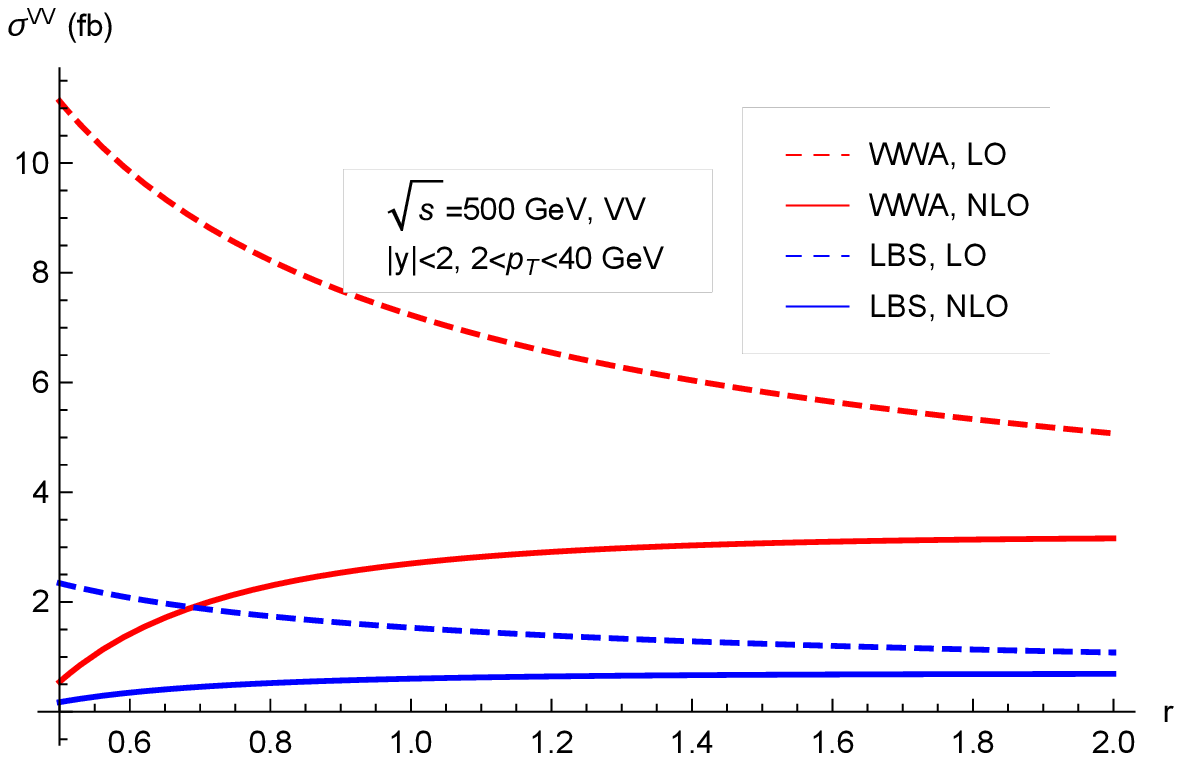}}
\caption{The LO and NLO cross sections versus $r$, where $r=\tfrac{\mu}{\sqrt{m_{B_c}^2+p_T^2}}$. (a) PP, 250 GeV; (b) PP, 500 GeV; (c) PV, 250 GeV; (d) PV, 500 GeV; (e) VV, 250 GeV; and (f) VV, 500 GeV.}
\label{fig_rdis}
\end{figure}

\begin{figure}[!htbp]
\centering
\subfigure[]{
\includegraphics[width=0.46\textwidth]{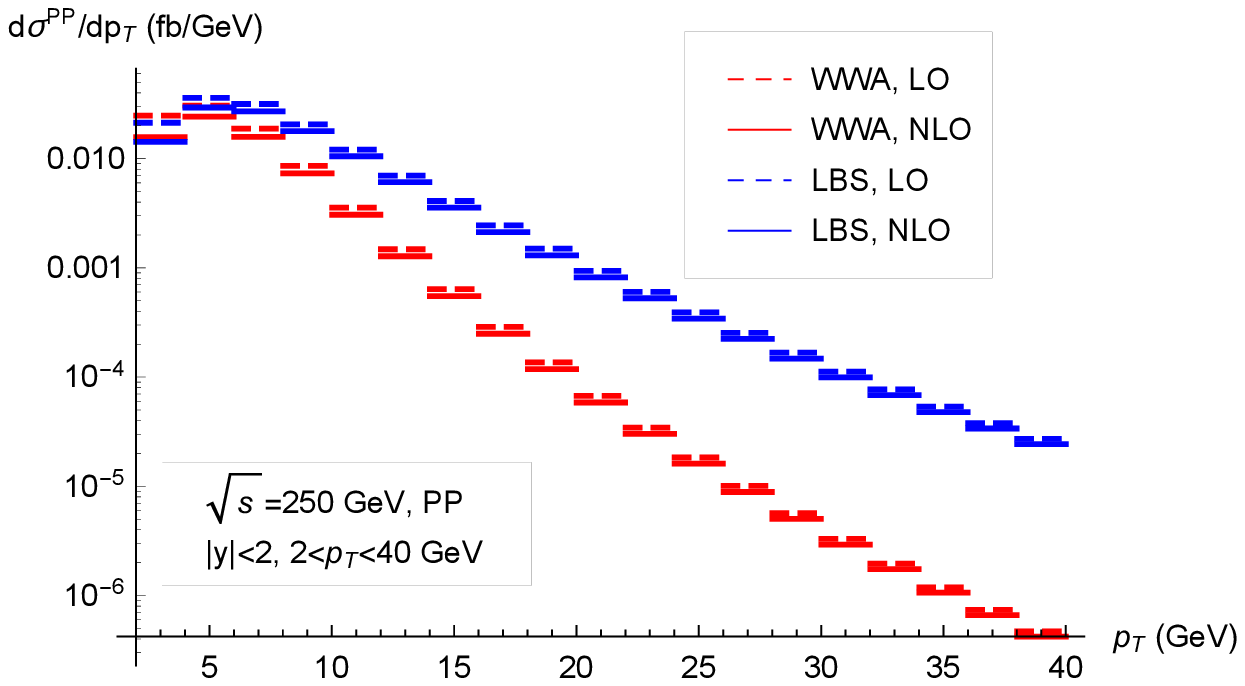}}
\subfigure[]{
\includegraphics[width=0.46\textwidth]{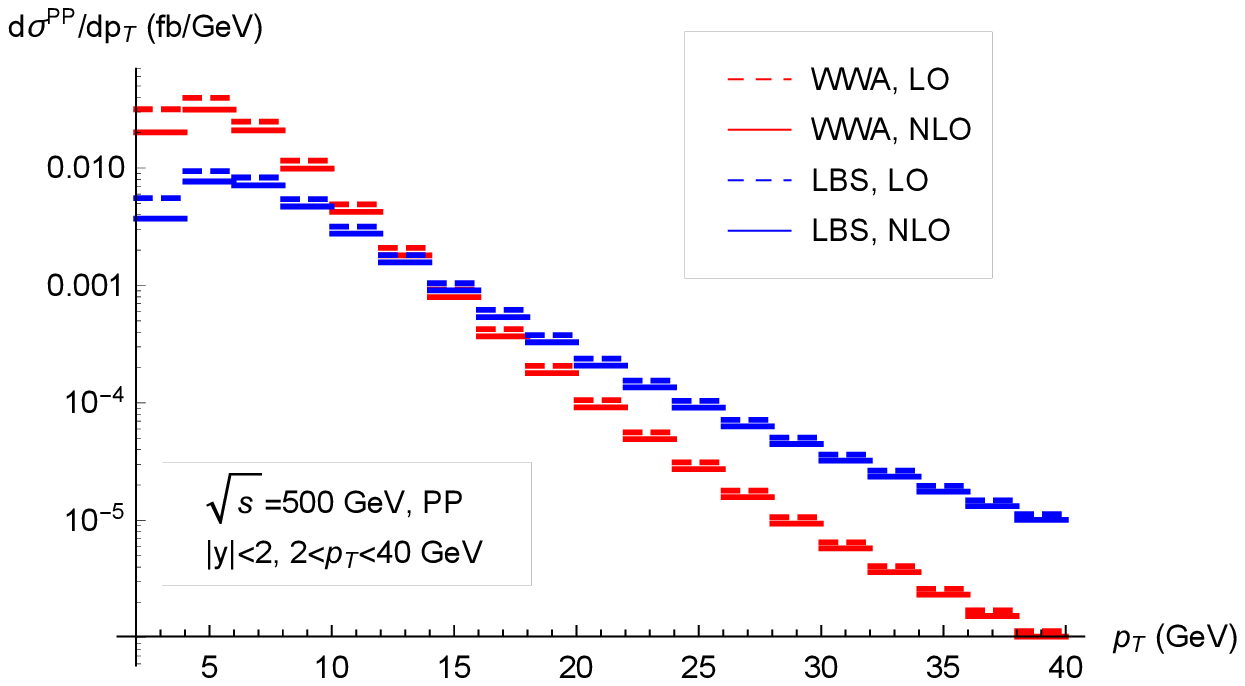}}
\subfigure[]{
\includegraphics[width=0.46\textwidth]{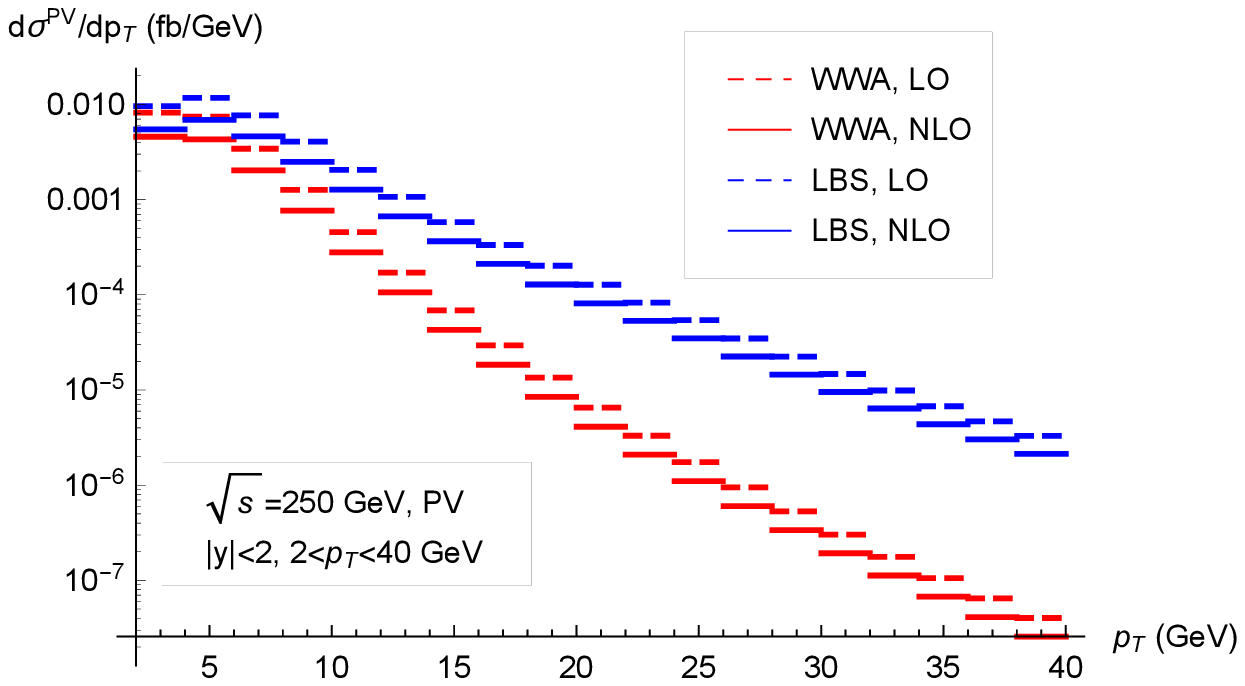}}
\subfigure[]{
\includegraphics[width=0.46\textwidth]{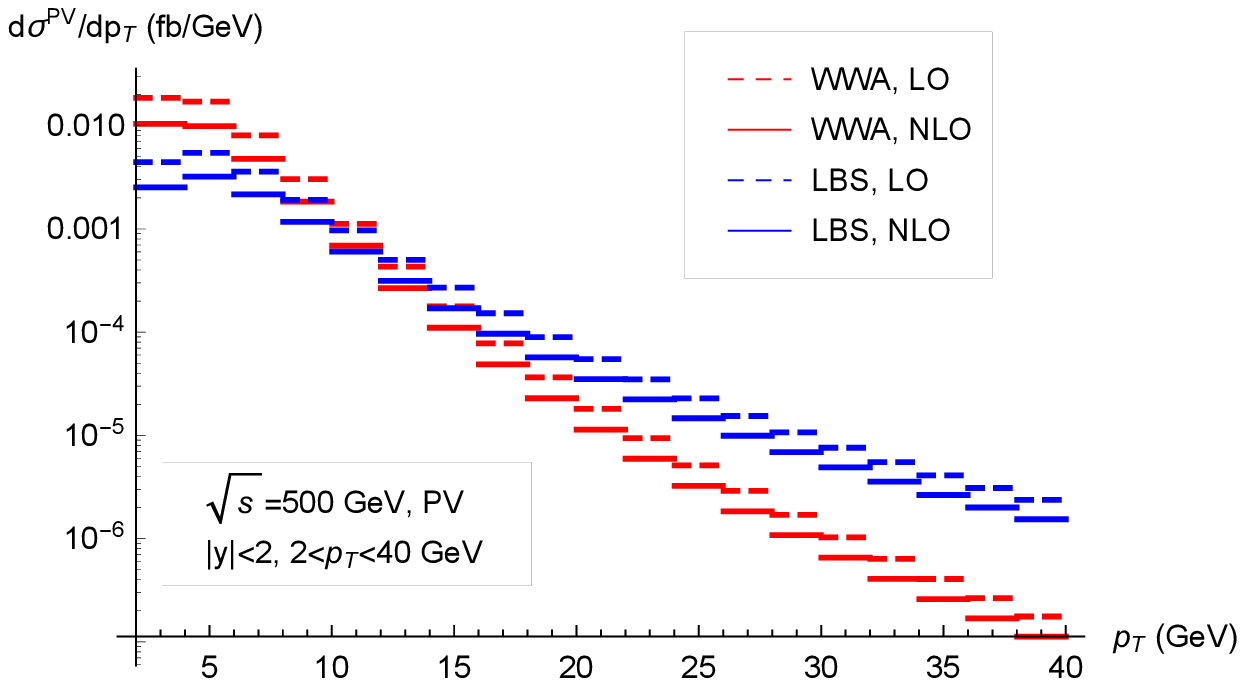}}
\subfigure[]{
\includegraphics[width=0.46\textwidth]{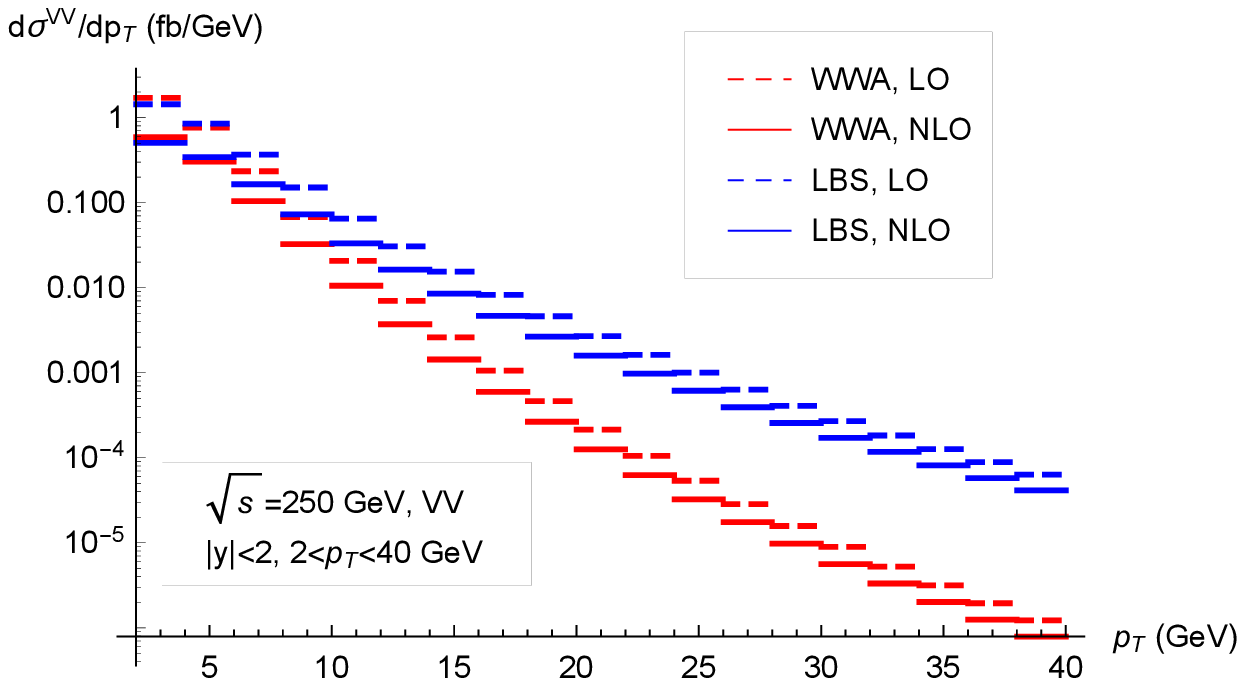}}
\subfigure[]{
\includegraphics[width=0.46\textwidth]{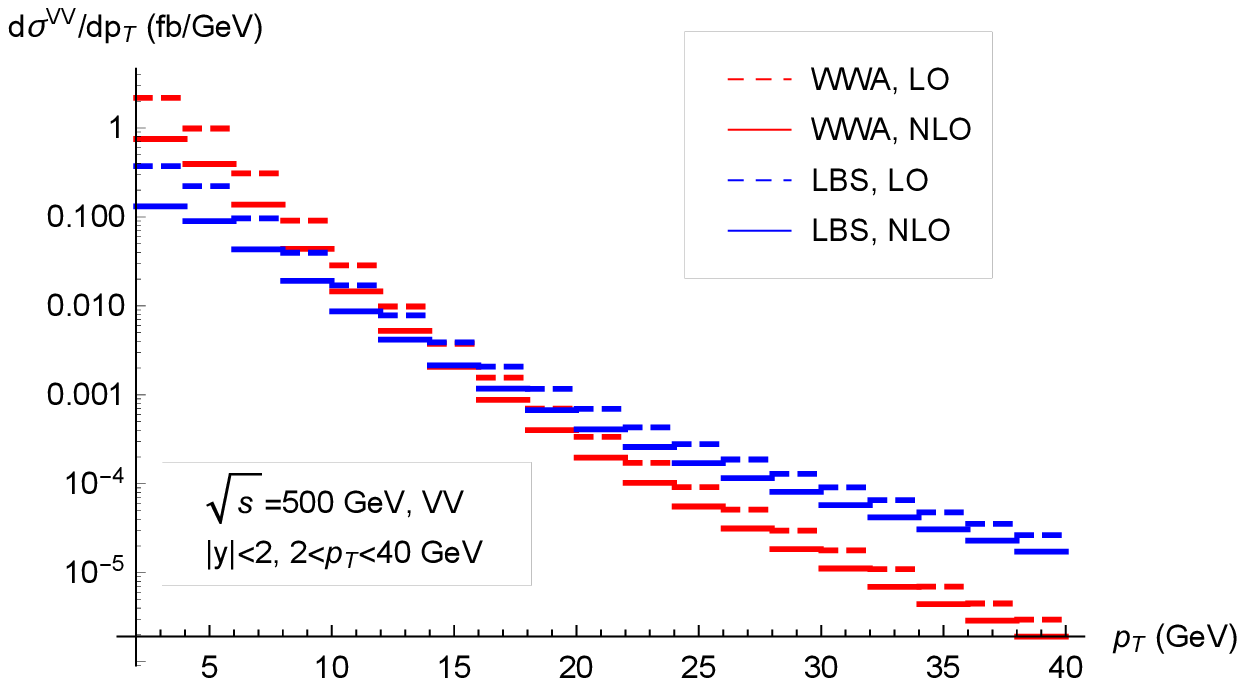}}
\caption{The LO and NLO differential cross sections versus $p_T$, the transverse momentum of one of the two $B_c$ mesons. The renormalization scale $\mu=\sqrt{m_{B_c}^2+p_T^2}$. (a) PP, 250 GeV; (b) PP, 500 GeV; (c) PV, 250 GeV; (d) PV, 500 GeV; (e) VV, 250 GeV; and (f) VV, 500 GeV.}
\label{fig_ptdis}
\end{figure}

\begin{figure}[!htbp]
\centering
\subfigure[]{
\includegraphics[width=0.46\textwidth]{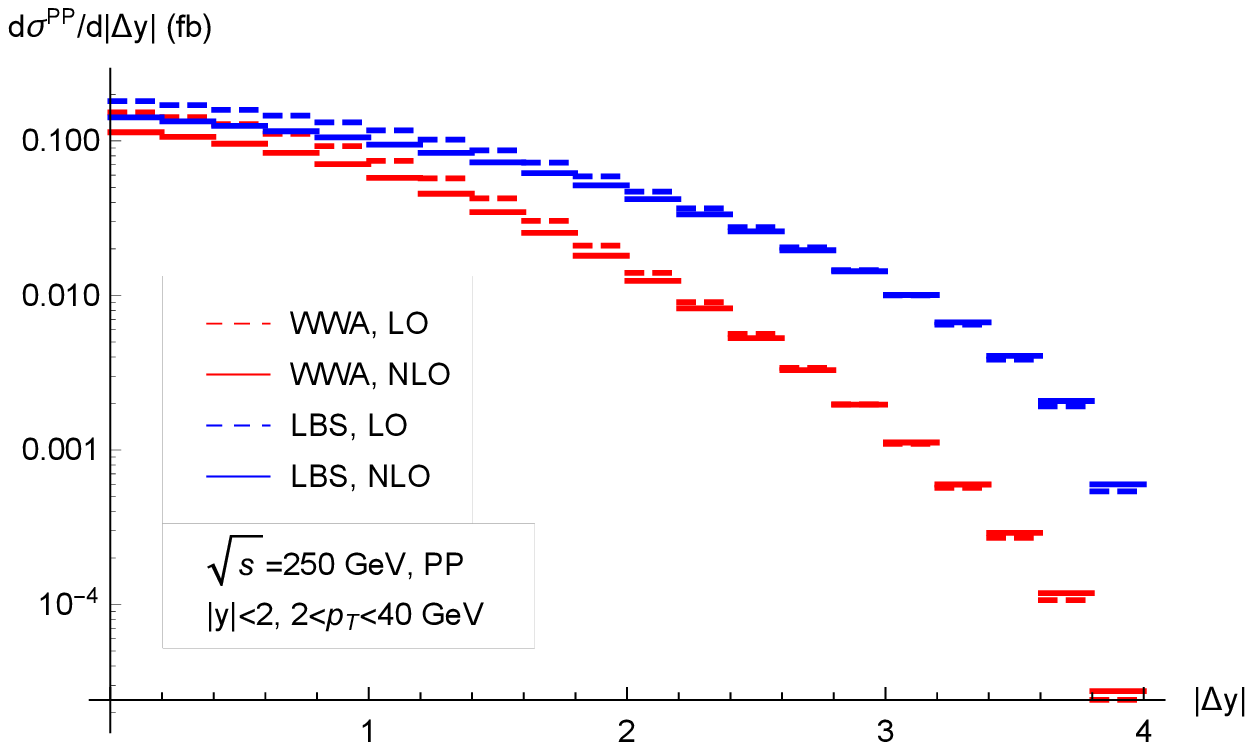}}
\subfigure[]{
\includegraphics[width=0.46\textwidth]{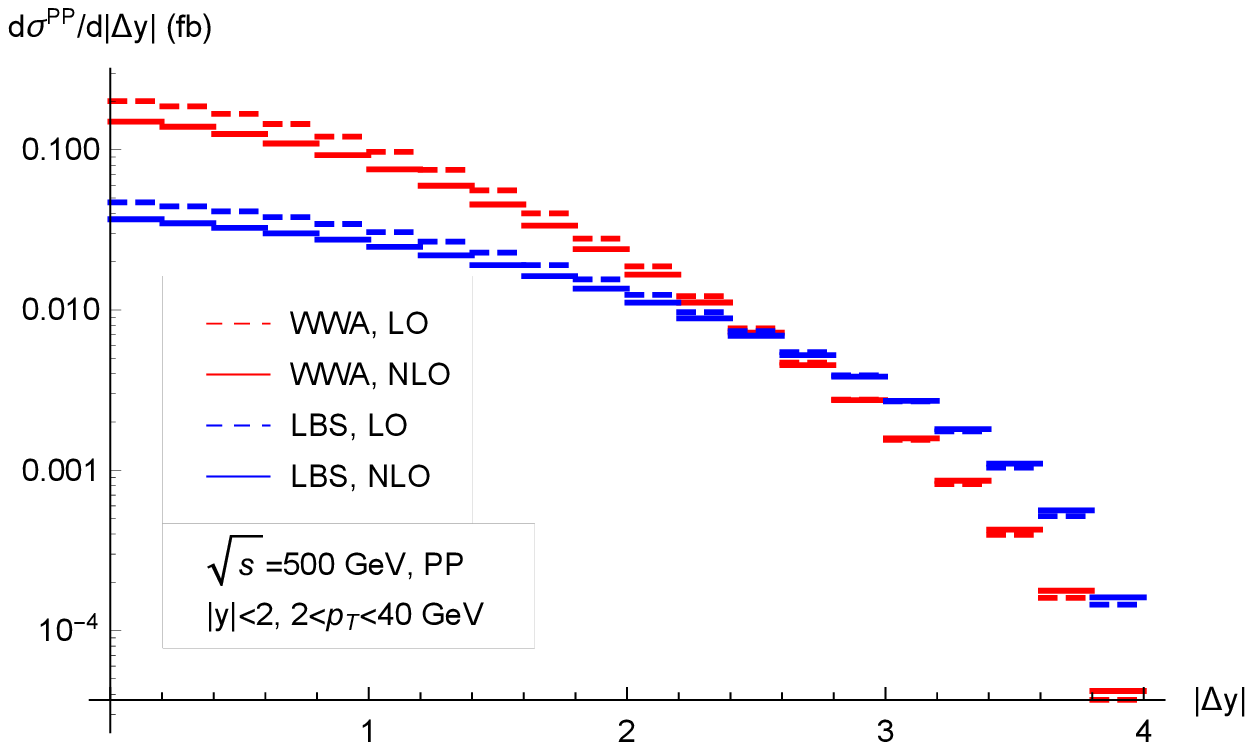}}
\subfigure[]{
\includegraphics[width=0.46\textwidth]{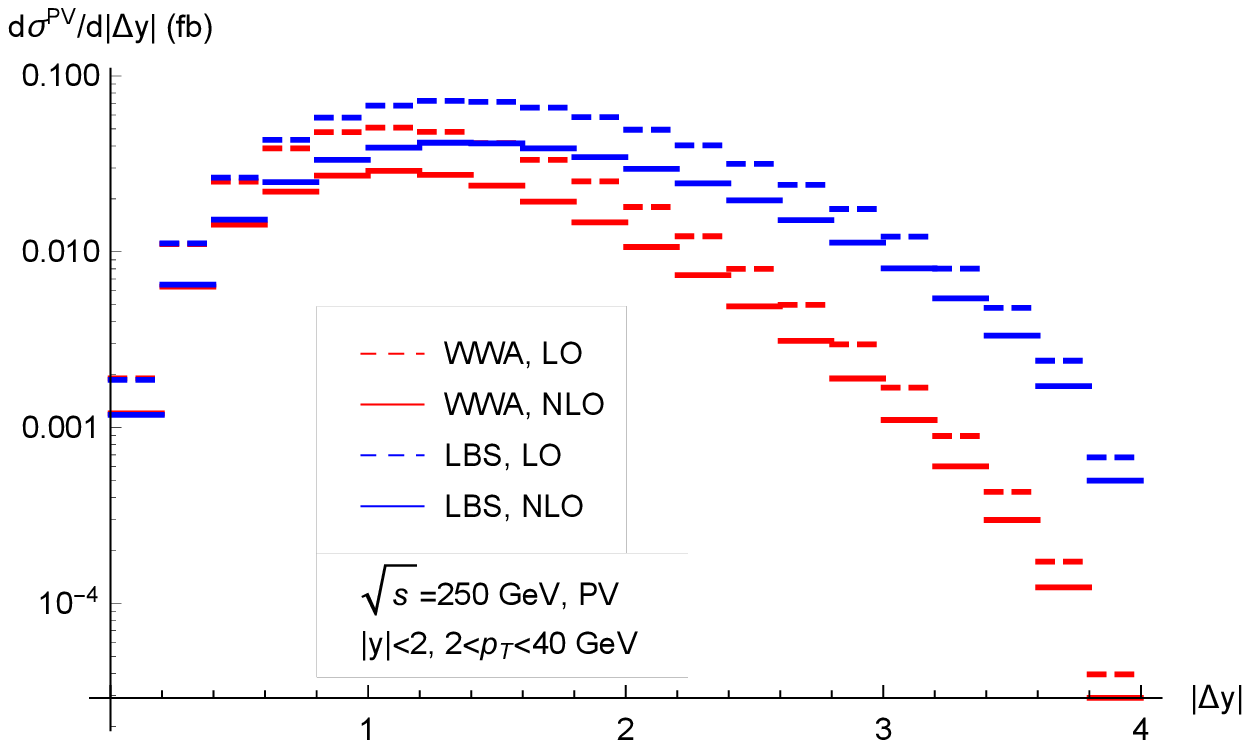}}
\subfigure[]{
\includegraphics[width=0.46\textwidth]{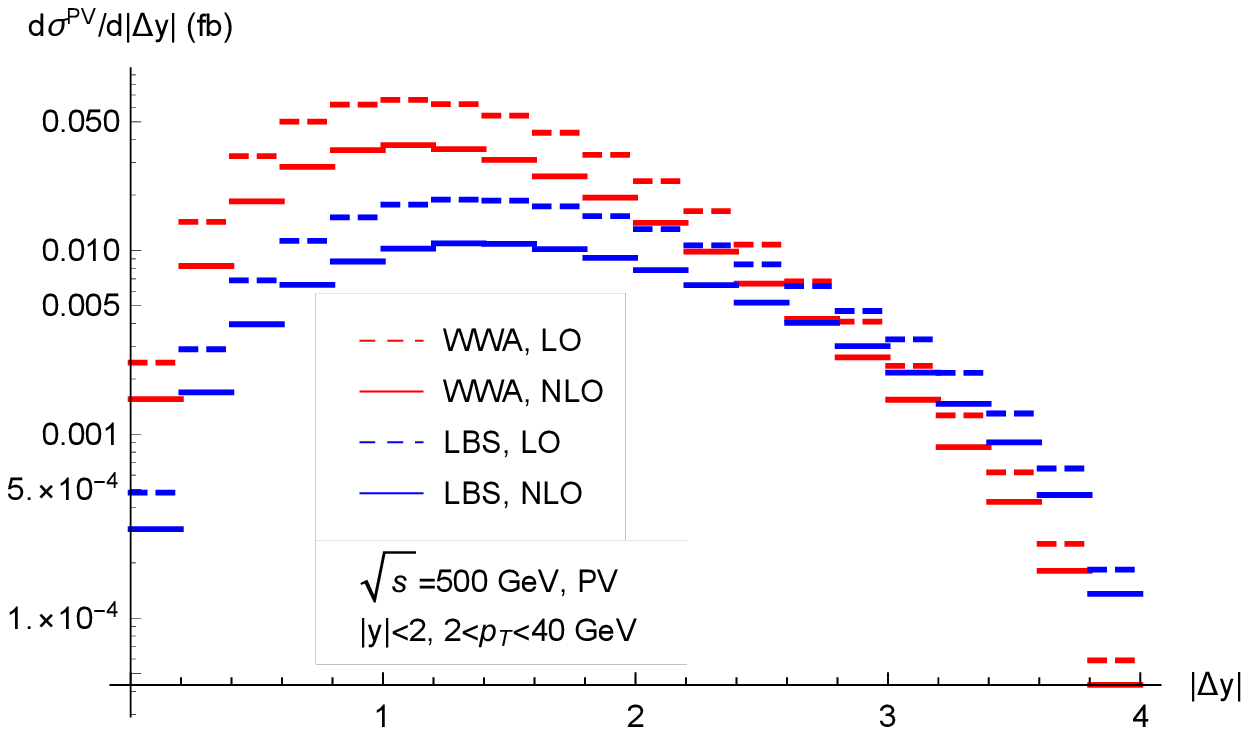}}
\subfigure[]{
\includegraphics[width=0.46\textwidth]{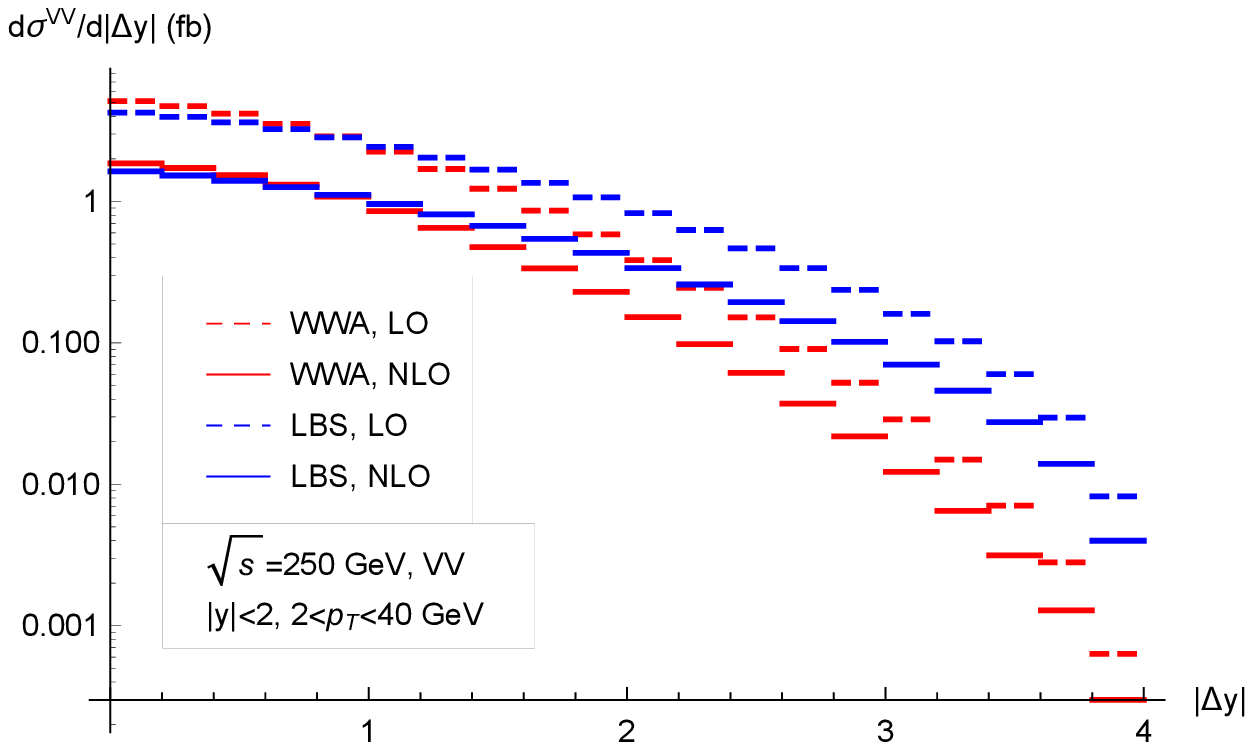}}
\subfigure[]{
\includegraphics[width=0.46\textwidth]{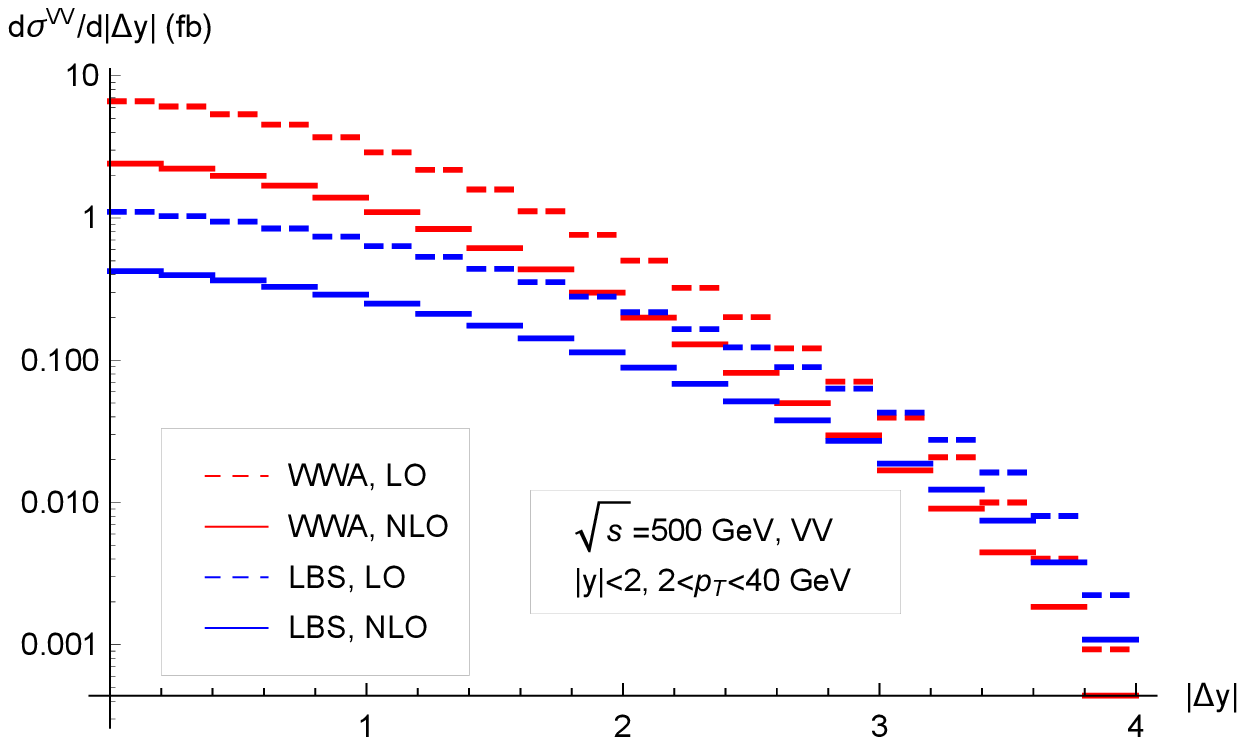}}
\caption{The LO and NLO differential cross sections versus $\Delta y$, the rapidity difference between the two $B_c$ mesons. The renormalization scale $\mu=\sqrt{m_{B_c}^2+p_T^2}$. (a) PP, 250 GeV; (b) PP, 500 GeV; (c) PV, 250 GeV; (d) PV, 500 GeV; (e) VV, 250 GeV; and (f) VV, 500 GeV.}
\label{fig_ysdis}
\end{figure}

Considering in future the $e^+e^-$ collider like CEPC and ILC might run at center-of-mass energies $\sqrt{s}=250$ GeV and $\sqrt{s}=500$ GeV respectively, we numerically evaluate the $B_c$-pair production via WWA  and LBS schemes at these two energies.
Taking the same inputs, we can numerically repeat the LO double pseudoscalar $B_c$ production result in \cite{Baranov:1997wy}. The full NLO results are presented in Fig.\ref{fig_rdis}, Fig.\ref{fig_ptdis} and Fig.\ref{fig_ysdis}.
Note, because the cross sections for PV production and VP production are exactly the same, only the PV production results are illustrated.

The total cross sections versus $r$ are shown in Fig.\ref{fig_rdis} with $\mu=r\sqrt{m_{B_c}^2+p_T^2}$.
We observe that, in comparison with the LO contribution, the LO plus NLO cross sections are suppressed, as are their dependences on the renormalization scale. As the $\sqrt{s}$ rises from 250 GeV to 500 GeV, the $B_c$-pair production rates increase in the WWA mechanism while decrease in the LBS mechanism.
This may be understood from the different behaviors of WWA and LBS mechanisms, as shown in Fig.\ref{fig_phdis}.
The WWA photons are more likely to be produced with small momentum fraction $x$, while the LBS photons tend to be more energetic.
Moreover, the partonic cross section $\hat{\sigma}(\gamma+\gamma\to B_c^++B_c^-)$ decreases with the increase of incident photons' center-of-mass energy.

The differential cross sections as functions of $p_T$, the transverse momentum of one of the two $B_c$ mesons, are shown in Fig.\ref{fig_ptdis}. It can be seen that as $\sqrt{s}$ increases from 250 GeV to 500 GeV, the yields of WWA processes increase slightly, while the yields of LBS processes decrease, evidently in small $p_T$ region.
Since the LBS photons are generally more energetic than the WWA photons, the produced $B_c$ pairs tend to have larger transverse momenta, which shall lead to a flatter $p_T$ distribution.

The differential cross sections as functions of $\Delta y$, the rapidity difference between two produced $B_c$ mesons, are shown in Fig.\ref{fig_ysdis}. Note, due to $|\Delta y|=2|y^*|$, the $|\Delta y|$ distribution is equivalent to the $|y^*|$ distribution, where $y^*$ is the rapidity of $B_c$ meson in the photon-photon center-of-mass frame.
For PP and VV production, the $B_c$ pairs are more likely to be produced around the $y^*=0$ region, while for the PV or VP production, the peak is located close to $y^*=0.6$. Since the large energy may lead to large $y*$, for the same reason as explained in $p_T$ distribution, the LBS production distributions are flatter than the WWA ones.

In Ref.\cite{Berezhnoy:2016etd}, the production of $B_c$ pairs in $e^+e^-$ annihilation via virtual $\gamma^*$ and $Z^*$ are investigated at the NLO QCD accuracy. At large $\sqrt{s}$, say $\sqrt{s}>160$ GeV, the cross sections are less than $10^{-6}$ fb, which are four to six orders of magnitude smaller than the cross sections of the processes considered here. It means that at high energy $e^+e^-$ collider, photon-photon collision turns out to be the dominant mechanism for $B_c$-pair production.

Moreover, about the numerical results, there are some points remarkable: \\
1) At $\sqrt{s}=$250 GeV, as can be seen from Fig.\ref{fig_phdis}, the WWA and LBS spectra meet at about $x\sim 0.07$, and hence
the WWA and LBS cross sections tend to be comparable in the corresponding kinematic region (i.e. in about $2<p_T<4$ GeV and $0<|\Delta y|<1$), as shown in Fig.\ref{fig_ptdis}(a)(c)(e) and Fig.\ref{fig_ysdis}(a)(c)(e).
For VV production, since small $p_T$ and small $|\Delta y|$ regions dominate the production, the WWA and LBS cross sections are hence almost identical, as shown in Fig.\ref{fig_rdis}(e). \\
2) To investigate the convergence of perturbative expansion, we define a measure $\mathcal{R}=|\frac{\sigma_{\rm NLO}-\sigma_{\rm LO}}{\sigma_{\rm LO}}|$.
For PP, PV and VV productions, we find
\begin{equation}
0<\mathcal{R}_{\rm PP}<0.5\ ,\quad 0.2<\mathcal{R}_{\rm PV}<0.7\ ,\quad 0.4<\mathcal{R}_{\rm VV}<0.9\ ,
\end{equation}
which are compatible with the results in Ref.\cite{Berezhnoy:2016etd}, where the $B_c$-pair production via $e^+e^-$ annihilation was studied. Furthermore, the NLO effect is more significant in the small $p_T$ region, suggests a resummation for the logarithms of $\frac{p_T^2}{\hat{s}}$, which is beyond the scope of this work.\\
3) In the numerical calculation, the strong coupling constant is parameterized by $\Lambda_{\rm QCD}$ as in Eq.(\ref{eq_alphasLambda}).
Whereas, it is noteworthy that nowadays an alternative approach is widely accepted, in which the initial value of $\alpha_s$ at a well experimentally measured point is adopted, usually at $M_Z$, rather than $\Lambda_{\rm QCD}$. Then, for fixed $n_f$, one may use evolution equation \cite{Deur:2016tte}
\begin{equation}
\frac{4\pi}{\alpha_s(\mu^2)}-\frac{\beta_1}{\beta_0}\ln\left(\frac{4\pi}{\alpha_s(\mu^2)}+\frac{\beta_1}{\beta_0}\right) =\frac{4\pi}{\alpha_s(M_Z^2)}-\frac{\beta_1}{\beta_0}\ln\left(\frac{4\pi}{\alpha_s(M_Z^2)} +\frac{\beta_1}{\beta_0}\right)+\beta_0\ln\frac{\mu^2}{M_Z^2}
\end{equation}
to run the $\alpha_s$ to where interested in. We evaluate the cross sections as well employing this parameterizing scheme with input $\alpha_s(M_Z^2)=0.1181$ \cite{Tanabashi:2018oca}, and find that in comparison with the results from the original scheme, the LO cross sections are generally suppressed substantially by a factor of about 0.7, while the dominance of the NLO results in two schemes alternates case to case, but with discrepancies less than $5\%$. As a good approximation, to match to the evolution scheme one may keep on using Eq.(\ref{eq_alphasLambda}), but with $\Lambda_{\rm QCD}$ determined by $\alpha_s(M_Z^2)$, i.e. $\Lambda_{\rm QCD}^{\rm LO}=88$ MeV, $\Lambda_{\rm QCD}^{\rm NLO}=228$ MeV. In all, the discrepancy in strong coupling constant parameterization between two schemes may be somehow remedied by adjusting the value of $\Lambda_{\rm QCD}$, but the evolution scheme is recommended.

\section{Summary}
In this work we investigated the $B_c$-pair production in high energy photon-photon fusion at the NLO accuracy in the NRQCD factorization framework. Various of $S$-wave $B_c$ states, including configurations of PP, PV, VP and VV, were taken into account. Considering the leading order results for $B_c$-pair production in PV and VP configurations are still missing in the literature, we calculated them and provided the analytic results. The total cross section as well as $p_T$ and $\Delta y$ distributions in $e^+e^-$ collider with $\sqrt{s}=$250 GeV and $\sqrt{s}=$ 500 GeV were evaluated and presented in figures.

The numerical results show that with the NLO corrections, the LO cross sections are suppressed, and their dependence on  renormalization scale are reduced evidently. By comparing with the results in Ref.\cite{Berezhnoy:2016etd}, where the $B_c$-pair production in $e^+e^-$ annihilation via virtual $\gamma^*$ and $Z^*$ were investigated, we may conclude that at large $e^+e^-$ collision energy, say $\sqrt{s}>160$ GeV, photon-photon collision will be the dominant source of $B_c$-pair production.

The NLO calculation of the concerned processes is somewhat time consuming and computer resource exhausting.
To fulfill this work, a "divide-and-conquer" strategy was employed. Instead of squaring the amplitude and summing over spins, we calculated the helicity amplitudes separately, which makes this tedious calculation workable.
Moreover, it shows that the symmetries remain in the helicity amplitudes may greatly reduce the number of independent amplitudes, as illustrated in Appendix. 

Last, the concerned processes involve a number of momenta and polarization vectors of the external particles, by introducing auxiliary vectors, the base $n_0$, $n_1$, $n_2$ and $n_3$, the number of independent Lorentz vectors reduces to 4, which facilitate the computation of Feynman integrals. We think the technical strategy employed in this work might be applicable to the studies of some other relevant processes.

Note added: when this work was finished and the manuscript was finalizing, there appeared a study on the web about the $B_c$-pair production in photon-photon collison with the relativistic corrections \cite{Dorokhov:2020nvv}.

\newpage
\vspace{0.0cm} {\bf Acknowledgments} \vspace{-0.0cm}

This work was supported in part by the National Natural Science Foundation of China(NSFC) under the Grants 11975236 and 11635009.


\section*{Appendix}
For PP, PV (or VP) and VV production, there are 4, 12 and 36 helicity amplitudes respectively,
whereas, half of them are zero. The nonzero helicity amplitudes are
\begin{align}
&\mathcal{M}^{11}_{\rm PP},\ \mathcal{M}^{44}_{\rm PP},\ \mathcal{M}^{111}_{\rm PV,VP},\ \mathcal{M}^{122}_{\rm PV,VP},\ \mathcal{M}^{123}_{\rm PV,VP},\  \mathcal{M}^{212}_{\rm PV,VP},\ \mathcal{M}^{213}_{\rm PV,VP},\ \mathcal{M}^{221}_{\rm PV,VP};\nonumber \\
&\mathcal{M}^{1111}_{\rm VV},\ \mathcal{M}^{1122}_{\rm VV},\ \mathcal{M}^{1123}_{\rm VV},\ \mathcal{M}^{1132}_{\rm VV},\ \mathcal{M}^{1133}_{\rm VV},\ \mathcal{M}^{1212}_{\rm VV},\ \mathcal{M}^{1213}_{\rm VV},\ \mathcal{M}^{1221}_{\rm VV},\ \mathcal{M}^{1231}_{\rm VV},\nonumber \\
&\mathcal{M}^{2112}_{\rm VV},\ \mathcal{M}^{2113}_{\rm VV},\ \mathcal{M}^{2121}_{\rm VV},\ \mathcal{M}^{2131}_{\rm VV},\ \mathcal{M}^{2211}_{\rm VV},\ \mathcal{M}^{2222}_{\rm VV},\ \mathcal{M}^{2223}_{\rm VV},\ \mathcal{M}^{2232}_{\rm VV},\ \mathcal{M}^{2233}_{\rm VV}.
\end{align}

The processes of PV and VP productions are correlated in charge-conjugation transformation, their cross sections should be exactly the same. According convention (\ref{eq_momchoice}) and (\ref{eq_polchoice}), the amplitudes satisfy
\begin{equation}
\mathcal{M}^{ijk}_{\rm PV}=\pm \mathcal{M}^{ijk}_{\rm VP},
\end{equation}
where the plus sign corresponds to $\{i,j,k\}=\{1,1,1\}$ and $\{2,2,1\}$, the minus sign corresponds to other cases.
In addition, the helicity amplitudes satisfy also
\begin{align}
&\mathcal{M}^{122}_{\rm PV}=\mathcal{M}^{212}_{\rm PV}\big|_{k_z\to -k_z},\ \mathcal{M}^{123}_{\rm PV}=-\mathcal{M}^{213}_{\rm PV}\big|_{k_z\to -k_z};\nonumber \\
&\mathcal{M}^{122}_{\rm VP}=\mathcal{M}^{212}_{\rm VP}\big|_{k_z\to -k_z},\ \mathcal{M}^{123}_{\rm VP}=-\mathcal{M}^{213}_{\rm VP}\big|_{k_z\to -k_z};\nonumber \\
&\mathcal{M}^{1123}_{\rm VV}=-\mathcal{M}^{1132}_{\rm VV},\ \mathcal{M}^{2223}_{\rm VV}=-\mathcal{M}^{2232}_{\rm VV},\nonumber \\
&\mathcal{M}^{1212}_{\rm VV}=-\mathcal{M}^{1221}_{\rm VV}\big|_{k_z\to -k_z}=-\mathcal{M}^{2112}_{\rm VV}\big|_{k_z\to -k_z}=\mathcal{M}^{2121}_{\rm VV},\nonumber \\
&\mathcal{M}^{1213}_{\rm VV}=-\mathcal{M}^{1231}_{\rm VV}\big|_{k_z\to -k_z}=\mathcal{M}^{2113}_{\rm VV}\big|_{k_z\to -k_z}=-\mathcal{M}^{2131}_{\rm VV}.
\end{align}

The analytical expressions for helicity amplitudes can be classified in photon-quark coupling, as
\begin{equation}
\mathcal{M}=\frac{8C_AC_Fm_{B_c}^3\pi^2\alpha\alpha_s}{3E_1^2m_b^2m_c^2}\left[e_c^2  f_1-e_ce_b(f_2+f_3)+e_b^2f_4+\sum_{i=u,d,s} e_i^2 f_5\right],
\end{equation}
where $e_q$ represents the electric charge number of quark $q$, i.e. $e_c=e_u=\frac{2}{3}$, $e_d=e_s=e_b=-\frac{1}{3}$.
The coefficients $f_1$ and $f_4$, $f_2$ and $f_3$ are related as per $m_c\leftrightarrow m_b$ exchange:
\begin{align}
&f_{\rm PP,1}\big|_{m_c\leftrightarrow m_b}=f_{\rm PP,4},\ f_{\rm PP,2}\big|_{m_c\leftrightarrow m_b}=f_{\rm PP,3}; \nonumber \\
&f_{\rm PV,1}\big|_{m_c\leftrightarrow m_b}=-f_{\rm PV,4},\ f_{\rm PV,2}\big|_{m_c\leftrightarrow m_b}=-f_{\rm PV,3}; \nonumber \\
&f_{\rm VP,1}\big|_{m_c\leftrightarrow m_b}=-f_{\rm VP,4},\ f_{\rm VP,2}\big|_{m_c\leftrightarrow m_b}=-f_{\rm VP,3}; \nonumber \\
&f_{\rm VV,1}\big|_{m_c\leftrightarrow m_b}=f_{\rm VV,4},\ f_{\rm VV,2}\big|_{m_c\leftrightarrow m_b}=f_{\rm VV,3}.
\end{align}

For the tree amplitudes, $f_5$ is zero. The analytical results for other coefficients are
\begin{align}
&f^{11}_{\rm PP,4}=-\tfrac{r^2 (1-r_z^2)}{r-1}-\tfrac{r_y^2}{(r-1) (1-r_z^2)}+\tfrac{r (r r_y^2-2 r+3)}{r-1}-\tfrac{2 r_y^2}{(1-r_z^2)^2},\nonumber \\
&f^{11}_{\rm PP,3}=1-\tfrac{2 r_y^2}{(1-r_z^2)^2},\nonumber \\
&f^{22}_{\rm PP,4}=-\tfrac{2 r_y^2 (2 r^2 r_y^2-4 r^2+4 r-1)}{(1-r_z^2)^2}-\tfrac{r^2 (1-r_z^2)}{r-1}+\tfrac{r_y^2 (4 r^3-2 r^2 r_y^2-6 r+1)}{(r-1) (1-r_z^2)}+\tfrac{r (3 r r_y^2-2 r+3)}{r-1},\nonumber \\
&f^{22}_{\rm PP,3}=\tfrac{2 (2 r^2-2 r-1) r_y^2}{1-r_z^2}-\tfrac{2 r_y^2 (2 r^2 r_y^2-4 r^2-2 r r_y^2+4 r-1)}{(1-r_z^2)^2}+1; \nonumber \\
&f^{111}_{\rm PV,4}=ir_mr_yr_z\big(\tfrac{r}{(r-1) (1-r_z^2)}+\tfrac{2}{(1-r_z^2)^2}\big),\nonumber \\
&f^{111}_{\rm PV,3}=-\tfrac{i2r_mr_yr_z}{(1-r_z^2)^2},\nonumber \\
&f^{122}_{\rm PV,4}=\tfrac{ir_mr_y}{\sqrt{r_y^2+r_z^2}}\big(\tfrac{2 r^2+r r_y^2-3 r+2}{(r-1) (1-r_z^2)}+\tfrac{2 (r r_y^2+r r_z-r+1)}{(1-r_z^2)^2}-\tfrac{r}{r-1}\big),\nonumber \\
&f^{122}_{\rm PV,3}=\tfrac{ir_mr_y}{\sqrt{r_y^2+r_z^2}}\big(\tfrac{2 (r r_y^2+r r_z-r-r_y^2-r_z)}{(1-r_z^2)^2}+\tfrac{2 r}{1-r_z^2}\big),\nonumber \\
&f^{123}_{\rm PV,4}=\tfrac{i}{\sqrt{r_y^2+r_z^2}}\big(-\tfrac{r r_y^2 (2 r-r_y^2-3)}{(r-1) (1-r_z^2)}+\tfrac{2 r_y^2 (r r_y^2+2 r r_z-r_z)}{(1-r_z^2)^2}-\tfrac{r (2 r_y^2+1)}{r-1}+\tfrac{r (1-r_z^2)}{r-1}\big),\nonumber \\
&f^{123}_{\rm PV,3}=\tfrac{i}{\sqrt{r_y^2+r_z^2}}\big(\tfrac{2 (r r_y^2+2 r r_z-r_y^2-r_z)}{(1-r_z^2)^2}-\tfrac{2 (r-1)}{1-r_z^2}\big),\nonumber \\
&f^{221}_{\rm PV,4}=ir_mr_yr_z\big(\tfrac{2 (2 r-1)}{(1-r_z^2)^2}-\tfrac{r}{(r-1) (1-r_z^2)}\big),\nonumber \\
&f^{221}_{\rm PV,3}=i2r_mr_yr_z\tfrac{-1+2r}{(1-r_z^2)^2};\nonumber \\
&f^{1111}_{\rm VV,4}=\tfrac{r^2 (1-r_z^2)}{r-1}-\tfrac{2 r-r_y^2-2}{(r-1) (1-r_z^2)}-\tfrac{r (r r_y^2+1)}{r-1}+\tfrac{2 r_y^2}{(1-r_z^2)^2},\nonumber \\
&f^{1111}_{\rm VV,3}=-\tfrac{2 r_y^2}{(1-r_z^2)^2}+\tfrac{2}{1-r_z^2}-1,\nonumber \\
&f^{1122}_{\rm VV,4}=\tfrac{1}{r_y^2+r_z^2}\big(-\tfrac{r^2 r_y^4+r^2 r_y^2+r r_y^2+3 r+r_y^2-2}{r-1}-\tfrac{r^2 (1-r_z^2)^2}{r-1}+\tfrac{r (1-r_z^2) (2 r r_y^2+r+1)}{r-1}\nonumber \\
&+\tfrac{(r_y^2+1) (2 r+r_y^2-2)}{(r-1) (1-r_z^2)}+\tfrac{2 (r_y-1) (r_y+1) r_y^2}{(1-r_z^2)^2}\big),\nonumber \\
&f^{1122}_{\rm VV,3}=\tfrac{1}{r_y^2+r_z^2}\big(-\tfrac{2 r_y^2}{(1-r_z^2)^2}+\tfrac{2 (2 r_y^2+1)}{1-r_z^2}-r_y^2-r_z^2-2\big),\nonumber \\
&f^{1123}_{\rm VV,4}=\tfrac{r_mr_yr_z}{r_y^2+r_z^2}\big(\tfrac{r (r_y^2+1)}{(r-1) (1-r_z^2)}-\tfrac{r}{r-1}+\tfrac{2 (r_y-1) (r_y+1)}{(1-r_z^2)^2}\big),\nonumber \\
&f^{1123}_{\rm VV,3}=\tfrac{r_mr_yr_z}{r_y^2+r_z^2}\big(\tfrac{2}{1-r_z^2}-\tfrac{2}{(1-r_z^2)^2}\big),\nonumber \\
&f^{1133}_{\rm VV,4}=\tfrac{1}{r_y^2+r_z^2}\big(-\tfrac{r^2 r_y^4+3 r^2 r_y^2+2 r^2-r r_y^2-r+r_y^2}{r-1}-\tfrac{r^2 (1-r_z^2)^2}{r-1}+\tfrac{(r_y^2+1) r_y^2}{(r-1) (1-r_z^2)}\nonumber \\
&+\tfrac{r (1-r_z^2) (2 r r_y^2+3 r-1)}{r-1}+\tfrac{2 (r_y-1) (r_y+1) r_y^2}{(1-r_z^2)^2}\big),\nonumber \\
&f^{1133}_{\rm VV,3}=\tfrac{1}{r_y^2+r_z^2}\big(\tfrac{2 r_y^2}{1-r_z^2}-\tfrac{2 r_y^2}{(1-r_z^2)^2}-r_y^2-r_z^2\big),\nonumber \\
&f^{1212}_{\rm VV,4}=\tfrac{r_m^2}{\sqrt{r_y^2+r_z^2}}\big(-\tfrac{r (r_y^2+1)}{(r-1) (1-r_z^2)}-\tfrac{2 (r r_y^2+r_z)}{(1-r_z^2)^2}+\tfrac{r}{r-1}\big),\nonumber \\
&f^{1212}_{\rm VV,3}=\tfrac{r_m^2}{\sqrt{r_y^2+r_z^2}}\big(-\tfrac{2 (r r_y^2-r_y^2-1)}{(1-r_z^2)^2}-\tfrac{2}{1-r_z^2}\big),\nonumber \\
&f^{1213}_{\rm VV,4}=\tfrac{r_mr_y}{\sqrt{r_y^2+r_z^2}}\big(\tfrac{r (2 r-r_y^2-3)}{(r-1) (1-r_z^2)}-\tfrac{2 (r r_y^2+r r_z+r-1)}{(1-r_z^2)^2}+\tfrac{r}{r-1}\big),\nonumber \\
&f^{1213}_{\rm VV,3}=\tfrac{r_mr_y}{\sqrt{r_y^2+r_z^2}}\big(\tfrac{2 (r-1)}{1-r_z^2}-\tfrac{2 (r r_y^2+r r_z+r-r_y^2-1)}{(1-r_z^2)^2}\big),\nonumber \\
&f^{2211}_{\rm VV,4}=\tfrac{2 r_y^2 (2 r^2 r_y^2-1)}{(1-r_z^2)^2}+\tfrac{r^2 (1-r_z^2)}{r-1}-\tfrac{4 r^3 r_y^2-2 r^2 r_y^4-4 r^2 r_y^2-2 r r_y^2-2 r+r_y^2+2}{(r-1) (1-r_z^2)}-\tfrac{r (3 r r_y^2+1)}{r-1},\nonumber \\
&f^{2211}_{\rm VV,3}=\tfrac{2 r_y^2 (2 r^2 r_y^2-2 r r_y^2-1)}{(1-r_z^2)^2}-\tfrac{2 (2 r^2 r_y^2-2 r r_y^2-r_y^2-1)}{1-r_z^2}-1,\nonumber \\
&f^{2222}_{\rm VV,4}=\tfrac{1}{r_y^2+r_z^2}\big(\tfrac{2 r_y^2 (2 r^2 r_y^4+2 r^2 r_y^2-r_y^2+1)}{(1-r_z^2)^2}-\tfrac{r^2 (1-r_z^2)^2}{r-1}+\tfrac{4 r^3 r_y^2-5 r^2 r_y^4-7 r^2 r_y^2-3 r r_y^2+r+r_y^2-2}{r-1}\nonumber \\
&-\tfrac{8 r^3 r_y^4+4 r^3 r_y^2-2 r^2 r_y^6-10 r^2 r_y^4-4 r^2 r_y^2-2 r r_y^4+2 r+r_y^4-r_y^2-2}{(r-1) (1-r_z^2)}+\tfrac{r (1-r_z^2) (4 r r_y^2+r+1)}{r-1}\big),\nonumber \\
&f^{2222}_{\rm VV,3}=\tfrac{1}{r_y^2+r_z^2}\big(\tfrac{2 r_y^2 (2 r^2 r_y^4+2 r^2 r_y^2-2 r r_y^4-2 r r_y^2-2 r_y^2-1)}{(1-r_z^2)^2}-\tfrac{2 (4 r^2 r_y^4+2 r^2 r_y^2-4 r r_y^4-2 r r_y^2-r_y^4-3 r_y^2-1)}{1-r_z^2}\nonumber \\
&4 r^2 r_y^2+-4 r r_y^2-3 r_y^2-r_z^2-2\big),\nonumber \\
&f^{2223}_{\rm VV,4}=\tfrac{r_mr_yr_z}{r_y^2+r_z^2}\big(\tfrac{r (4 r-r_y^2-5)}{(r-1) (1-r_z^2)}-\tfrac{2 (2 r r_y^2+2 r+r_y^2-1)}{(1-r_z^2)^2}+\tfrac{r}{r-1}\big),\nonumber \\
&f^{2223}_{\rm VV,3}=\tfrac{r_mr_yr_z}{\sqrt{r_y^2+r_z^2}}\big(\tfrac{2 (2 r-1)}{1-r_z^2}-\tfrac{2 (2 r r_y^2+2 r+r_y^2 r_z-r_y^2-1)}{(1-r_z^2)^2}\big),\nonumber \\
&f^{2233}_{\rm VV,4}=\tfrac{1}{r_y^2+r_z^2}\big(\tfrac{2 r_y^2 (2 r^2 r_y^4+6 r^2 r_y^2+4 r^2-4 r r_y^2-4 r-r_y^2+1)}{(1-r_z^2)^2}-\tfrac{r^2 (1-r_z^2)^2}{r-1}+\tfrac{r (1-r_z^2) (4 r r_y^2+3 r-1)}{r-1}\nonumber \\
&-\tfrac{r_y^2 (8 r^3 r_y^2+12 r^3-2 r^2 r_y^4-14 r^2 r_y^2-24 r^2+2 r r_y^2+10 r+r_y^2+1)}{(r-1) (1-r_z^2)}+\tfrac{4 r^3 r_y^2-5 r^2 r_y^4-13 r^2 r_y^2-2 r^2+3 r r_y^2+r+r_y^2}{r-1}\big),\nonumber \\
&f^{2233}_{\rm VV,3}=\tfrac{1}{r_y^2+r_z^2}\big(-\tfrac{2 r_y^2 (4 r^2 r_y^2+6 r^2-4 r r_y^2-6 r-r_y^2)}{1-r_z^2}+\tfrac{2 r_y^2 (2 r^2 r_y^4+6 r^2 r_y^2+4 r^2-2 r r_y^4-6 r r_y^2-4 r+1)}{(1-r_z^2)^2}\nonumber \\
&+4 r^2 r_y^2-4 r r_y^2-3 r_y^2-r_z^2\big).
\end{align}
Here, $r=m_b/(m_b+m_c)$.
The analytical results for the one-loop amplitude are lengthy, and are presented in the supplementary files attached to the arXiv preprint.
\end{document}